\documentstyle[12pt,aasms4]{article}

\begin{document}


\title{Perturbative Approach to the Astrometric Microlensing 
due to an Extrasolar Planet} 

\author{Hideki Asada\altaffilmark{1}}
\affil{ 
Faculty of Science and Technology, 
Hirosaki University, Hirosaki 036-8561, Japan
} 
\authoremail{asada@phys.hirosaki-u.ac.jp}
\altaffiltext{1}{Visiting Researcher, Institute for Astrophysics 
at Paris} 

\begin{abstract}
We have developed a perturbative approach to microlensing 
due to an extrasolar planetary lens. 
In particular, we have found analytic formulae for triple images. 
We have used the formulae to investigate the astrometric 
microlensing due to the extrasolar planetary lens, 
in expectation of dramatic improvements in the precision 
of the future astrometric measurements. 
For a weak lensing case, we have shown how the maximum angular size 
and the typical time scale of the anomalous shift of the light
centroid are dependent on the mass ratio and angular separation between 
the star and the planet. 
\end{abstract}


\section{Introduction}
Extrasolar planets searches are successfully going on
(Marcy \& Butler 1998). 
The number of candidates for the planets has reached about 70 
thanks to the Doppler method. 
In spite of the success, this technique has some limitations:  
The inferred mass is the lower bound, 
since the inclination angle of the orbital plane is not determined 
except for the eclipse case. 
In addition, the radial velocity of a star becomes too small 
to detect its Doppler effect when separation between the star 
and the planet is of the order of an AU. 
Hence, it seems quite difficult to discover earth-type planets 
by this method. 
Astrometry is considered as a supplementary method to find out 
planets by measuring the transverse motion of a star. 
Indeed, we can expect significant improvements in the precision 
of the future astrometric measurements by DIVA, FAME, GAIA and SIM. 
Their precision will achieve a few micro arcseconds. 

These future astrometry missions make the astrometric 
microlensing the third method for searching extrasolar planets: 
The typical scale in the gravitational lensing is the Einstein ring 
radius, which is of the order of an AU for a lensing star 
within our galaxy. 
Hence, the microlensing is quite effective even for earth-type 
planets (Mao \& Paczynski 1991, Gould \& Loeb 1992). 
In the planetary lens case, we can determine its true mass 
through observation of spikes in the light curve, which are produced 
by the magnification effect of the gravitational lensing. 
On the other hand, the astrometric microlensing is a consequence 
of a combination of the magnification and the position shift 
of the image (Miyamoto and Yoshii 1995). 
The numerical results (Safizadeh et al. 1999, Han and Lee 2002) 
have shown that the photo centroid shifts provide us a clue 
for extrasolar planets. 
However, the parameter dependence is not clear 
in the numerical approach. 
The purpose of this paper is to derive the 
analytic formulae and clarify the parameter dependence. 
Analytic approaches to a binary lens have not been developed 
until now. 
After presenting a brief summary of the astrometric microlensing 
due to a single lens, we treat a planetary lensing case as a
perturbation around the stellar lens.

\section{Microlensing due to a planetary system} 
\subsection{A single lens}
A single lens is located at the distance $D_L$ from the observer, 
and a source at $D_S$. 
The distance between the lens and the source is denoted by $D_{LS}$. 
Under the thin lens approximation, the lens equation 
for the single lens is written as 
\begin{equation}
{\mbox{\boldmath $\beta$}}={\mbox{\boldmath $\theta$}}
-\frac{D_{LS}}{D_S}{\mbox{\boldmath $\alpha$}} , 
\end{equation}
where {\boldmath $\beta$} and {\boldmath $\theta$} are 
the angular positions of the source and the image, respectively, 
and {\boldmath $\alpha$} is the deflection angle. 
For a spherical lens with mass $M$, or at least axisymmetric along the
line of sight, the deflection angle is 
\begin{equation}
|{\mbox{\boldmath $\alpha$}}|=\frac{4GM}{c^2 b} ,  
\end{equation}
where $G$ is the gravitational constant and $b$ is the impact 
parameter of the light ray. 
It is convenient to normalize our equations by the angular radius of 
the Einstein ring 
\begin{equation}
\theta_E=\sqrt{\frac{4GM D_{LS}}{c^2 D_L D_S}} ,  
\end{equation}
which is of the order of a milli arcsecond (mas) for a solar mass 
lens within our galaxy. 
In units of the Einstein ring radius, the lens equation is 
rewritten as 
\begin{equation}
{\mbox{\boldmath $\beta$}}={\mbox{\boldmath $\theta$}}
-\frac{{\mbox{\boldmath $\theta$}}}{\theta^2} , 
\end{equation}
where $\theta$ denotes $|\mbox{\boldmath $\theta$}|$ and 
similar notations are used below. 
We find out the two solutions 
\begin{equation}
{\mbox{\boldmath $\theta$}}^{(\pm)}=
k^{(\pm)} {\mbox{\boldmath $\beta$}} , 
\end{equation}
where we defined 
\begin{equation}
k^{(\pm)}=\frac12 \Bigl(1\pm\sqrt{1+\frac{4}{\beta^2}}\Bigr) . 
\end{equation}

The amplification for each image is respectively 
\begin{equation}
A^{(\pm)}=\frac12 \Bigl(\frac{\beta^2+2}
{\beta \sqrt{\beta^2+4}} \pm 1\Bigr) . 
\label{amplification}
\end{equation}
Following the definition of the center of mass, 
we define the photo center as (Walker 1995) 
\begin{equation}
{\mbox{\boldmath $\theta$}_C}=\frac{A^{(+)} 
\mbox{\boldmath $\theta$}^{(+)}+A^{(-)} 
\mbox{\boldmath $\theta$}^{(-)}}{A^{(+)}+A^{(-)}} . 
\end{equation}
With respect to the unlensed position, the location of 
the photo center is 
\begin{equation}
\Delta {\mbox{\boldmath $\theta$}_C}=
{\mbox{\boldmath $\theta$}_C}-{\mbox{\boldmath $\beta$}} ,  
\end{equation}
which expresses the deviation due to the lensing. 
For the single lens, it becomes simply 
\begin{equation}
\Delta {\mbox{\boldmath $\theta$}_C}=
\frac{\mbox{\boldmath $\beta$}}{\beta^2+2} , 
\label{ellipse}
\end{equation}
where we used Eq. ($\ref{amplification}$). 
We can assume that the stellar motion in our galaxy is 
approximated over several decades by a straight line, since 
the curvature of the orbit in our galaxy is negligible. 
For the later convenience, we decompose $\mbox{\boldmath $\beta$}$ 
into the orthogonal vectors as 
\begin{equation}
\mbox{\boldmath $\beta$}=\mbox{\boldmath $b$}+\mbox{\boldmath $V$} , 
\label{decomposition}
\end{equation}
where $\mbox{\boldmath $b$}$ is the vector for the impact parameter 
from the lens to the stellar orbit, and 
\begin{equation}
\mbox{\boldmath $V$}=t \mbox{\boldmath $v$}_{\perp} , 
\end{equation}
for $\mbox{\boldmath $v$}_{\perp}$, the transverse angular velocity 
of the source to the lens. 
Here, we have chosen $t=0$ as the time when the star is closest to  
the lens.  
We introduce $\xi$ as 
\begin{equation}
V=\sqrt{\beta^2+2}\tan\xi ,  
\end{equation}
so that the photo center can be rewritten as 
\begin{equation}
\Delta {\mbox{\boldmath $\theta$}_C}
-\frac{\mbox{\boldmath $b$}}{2 (b^2+2)} = 
\frac{\mbox{\boldmath $b$} \cos{2\xi}}{2 (b^2+2)}
+\frac{\mbox{\boldmath $V$}\sin{2\xi}}{2V \sqrt{b^2+2}} , 
\end{equation}
which is an ellipse (Walker 1995, Jeong et al. 1999). 
Its size is of the order of a mas for stellar cases in our galaxy. 
The center of the ellipse is at 
$\frac{\mbox{\boldmath $b$}}{2 (b^2+2)}$, 
and the length of the semimajor and semiminor axes are 
$\frac{1}{2\sqrt{b^2+2}}$ and $\frac{b}{2 (b^2+2)}$, respectively. 
The photo center passes the two points on the semiminor axes 
at $t=0 \quad(\xi=0)$ and $t=\pm\infty \quad(\xi=\pm\frac{\pi}{2})$.

\subsection{A planetary lens}
Now, we are in the position to consider the astrometric 
microlensing due to the planetary system, where the mass 
of the star and the planet are $M_1$ and $M_2$ respectively. 
Let us consider the separation vector from the star to the planet. 
Its projection onto the lens plane is denoted by 
$\mbox{\boldmath $s$}$. 
We adopt the frame of center of mass. 
Then, in the unit of the Einstein ring radius angle 
due to the total mass $M_1+M_2$, the lens equation is written as 
\begin{equation}
\mbox{\boldmath $\beta$}=\mbox{\boldmath $\theta$}
-\Bigl( 
\nu_1 \frac{\mbox{\boldmath $\theta$}+\nu_2 
\mbox{\boldmath $\epsilon$}}{|\mbox{\boldmath $\theta$}
+\nu_2 \mbox{\boldmath $\epsilon$}|^2} 
+\nu_2 \frac{\mbox{\boldmath $\theta$}-\nu_1 
\mbox{\boldmath $\epsilon$}}{|\mbox{\boldmath $\theta$}
-\nu_1 \mbox{\boldmath $\epsilon$}|^2} 
\Bigr) , 
\end{equation}
where we defined 
\begin{eqnarray}
\nu_1&=&\frac{M_1}{M_1+M_2} , \\
\nu_2&=&\frac{M_2}{M_1+M_2} , \\
\mbox{\boldmath $\epsilon$}&=&\frac{\mbox{\boldmath $s$}}{D_L} . 
\end{eqnarray}

In the planetary case, $M_2$ is much smaller than $M_1$. 
For instance, the Jupiter mass is about $10^{-3}$ 
of the solar mass. 
Hence, we introduce an expansion parameter as $\nu=\nu_2$ 
in our perturbation approach. 
Since we wish to consider the microlensing as a method 
supplementary to the Doppler technique, we concentrate 
on a large separation case $\epsilon > 1$, which 
is beyond the reach of the Doppler method. 
It is straightforward to extend our investigation to the case 
of $\epsilon < 1$. 
In addition, we consider a case of a large impact parameter 
$\beta \gg 1$, which is most probable because of 
the large cross section. 
In total, we consider the case of $\beta \gg \epsilon > 1$. 
Let us look for the solutions of the lens equation 
by taking a form of 
\begin{equation}
\mbox{\boldmath $\theta$}=\mbox{\boldmath $\theta$}_0
+\delta\mbox{\boldmath $\theta$} , 
\end{equation}
where $\mbox{\boldmath $\theta$}_0$ and 
$\delta\mbox{\boldmath $\theta$}$ are the solutions at the 
zeroth and first order, respectively. 

We know the zeroth-order solution 
\begin{equation}
\mbox{\boldmath $\theta$}_0^{(\pm)}=k^{(\pm)} 
{\mbox{\boldmath $\beta$}} , 
\end{equation}
which is expanded as 
\begin{eqnarray}
&&\theta_0^{(+)} \sim \beta \gg \epsilon , \\
&&\theta_0^{(-)} \sim \frac{1}{\beta} 
\ll \epsilon . 
\end{eqnarray}
That is, the ``$+$'' image appears outside the planetary system, 
while the ``$-$'' image inside it. 
Throughout our calculations, we should keep in mind that 
the large impact parameter does {\it not} mean 
$\theta \gg \epsilon$. 

Next, we find the solutions at the first order as 
\begin{eqnarray}
&&\delta\mbox{\boldmath $\theta$}^{(+)}=
\frac{\nu}{\beta^6} 
\Bigl( 
-2 {\mbox{\boldmath $\epsilon$}}({\mbox{\boldmath $\beta$}}\cdot
{\mbox{\boldmath $\epsilon$}}) \beta^2
-{\mbox{\boldmath $\beta$}}\epsilon^2 \beta^2
+4 {\mbox{\boldmath $\beta$}} ({\mbox{\boldmath $\beta$}}\cdot
{\mbox{\boldmath $\epsilon$}})^2
+O(\epsilon^3\beta^2) \Bigr) , 
\label{shift1}\\
&&\delta\mbox{\boldmath $\theta$}^{(-)}=\nu 
\Bigl( 
-{\mbox{\boldmath $\epsilon$}}
+\frac{\mbox{\boldmath $\beta$}}{\beta^2}
+O(\frac{1}{\epsilon\beta^2})
\Bigr) , 
\label{shift2}\\
&&\mbox{\boldmath $\theta$}^{(3)}=
{\mbox{\boldmath $\epsilon$}}-\nu 
\Bigl( {\mbox{\boldmath $\epsilon$}}
+\frac{{\mbox{\boldmath $\beta$}}+(\frac{1}{\epsilon^2}-1) 
{\mbox{\boldmath $\epsilon$}}}
{|{\mbox{\boldmath $\beta$}}+(\frac{1}{\epsilon^2}-1) 
{\mbox{\boldmath $\epsilon$}}|^2} \Bigr) , 
\label{shift3}
\end{eqnarray}
where the first and second solutions are perturbations around 
the zeroth-order solutions, and the third solution does not 
appear until at the linear order of $\nu$. 
Actually, for $\nu=0$, the third image is located at 
the direction to the planet. 
As for the number of the images, see also Witt (1993). 

Hence, we obtain the Jacobian of the mapping between 
the source and the lens planes, 
\begin{eqnarray}
&&|\frac{\partial{\mbox{\boldmath $\theta$}}^{(+)}}
{\partial{\mbox{\boldmath $\beta$}}}|=
1+O(\frac{\nu\epsilon^4}{\beta^6}, \nu^2) , 
\label{amplificationplus}\\
&&|\frac{\partial{\mbox{\boldmath $\theta$}}^{(-)}}
{\partial{\mbox{\boldmath $\beta$}}}|=
-\frac{1-2\nu}{\beta^4}+O(\nu^2) , \\
&&|\frac{\partial{\mbox{\boldmath $\theta$}}^{(3)}}
{\partial{\mbox{\boldmath $\beta$}}}|=O(\nu^2) . 
\label{jacobian3}
\end{eqnarray}
The contribution of the third solution to the photo center is
negligible for the large impact parameter. 

\subsection{Distortion of an ellipse for light centroid shifts}
By the use of the preceding subsection, we find a correction 
to the photo centroid as 
\begin{equation}
\delta{\mbox{\boldmath $\theta$}_C}=
\frac{\nu}{\beta^6} \Bigl( 
-2 {\mbox{\boldmath $\epsilon$}}({\mbox{\boldmath $\beta$}}\cdot
{\mbox{\boldmath $\epsilon$}}) \beta^2
-{\mbox{\boldmath $\beta$}}\epsilon^2 \beta^2
+4 {\mbox{\boldmath $\beta$}} ({\mbox{\boldmath $\beta$}}\cdot
{\mbox{\boldmath $\epsilon$}})^2
+O(\beta^3) \Bigr) +O(\nu^2) . 
\label{photoshift}
\end{equation}
This is due to a primary effect caused by a position shift 
given by Eq. ($\ref{shift1}$), since Eq. ($\ref{amplificationplus}$) 
shows that the correction to amplification appears 
only at higher orders. 

Let us investigate in detail the photo center shift. 
We consider the expansion of the following superposition of 
two ellipses 
\begin{eqnarray}
&&(1-\nu)
\frac{{\mbox{\boldmath $\beta$}}+\nu{\mbox{\boldmath $\epsilon$}}}
{|{\mbox{\boldmath $\beta$}}+\nu{\mbox{\boldmath $\epsilon$}}|^2 +2} 
+\nu\frac{{\mbox{\boldmath $\beta$}}-{\mbox{\boldmath $\epsilon$}}}
{|{\mbox{\boldmath $\beta$}}-{\mbox{\boldmath $\epsilon$}}|^2 +2}
\nonumber\\
&=&\frac{{\mbox{\boldmath $\beta$}}}{\beta^2 +2}
+\frac{\nu}{\beta^6} \Bigl( 
-2 {\mbox{\boldmath $\epsilon$}}({\mbox{\boldmath $\beta$}}\cdot
{\mbox{\boldmath $\epsilon$}}) \beta^2
-{\mbox{\boldmath $\beta$}}\epsilon^2 \beta^2
+4 {\mbox{\boldmath $\beta$}} ({\mbox{\boldmath $\beta$}}\cdot
{\mbox{\boldmath $\epsilon$}})^2
+O(\epsilon^3\beta^2) \Bigr)+O(\nu^2) , 
\end{eqnarray}
where the first term of the left hand side is considered 
as the primary ellipse and the second as the perturbation 
around the primary. 
The position of the first ellipse is shifted by 
$\nu{\mbox{\boldmath $\epsilon$}}$ in comparison with 
Eq. $(\ref{ellipse})$, and its size changes by factor $1-\nu$. 
Next, let us take a closer look at the secondary ellipse. 
The orthogonal decomposition by Eq. ($\ref{decomposition}$) is 
modified as 
\begin{equation}
{\mbox{\boldmath $\beta$}}-{\mbox{\boldmath $\epsilon$}}
=({\mbox{\boldmath $b$}}-{\mbox{\boldmath $\epsilon$}}_{\perp})
+(t-t_C){\mbox{\boldmath $v$}_{\perp}} , 
\end{equation}
where we defined 
\begin{eqnarray}
{\mbox{\boldmath $\epsilon$}}_{\parallel}&=&
\frac{({\mbox{\boldmath $\epsilon$}}
\cdot{\mbox{\boldmath $v$}_{\perp}}) 
{\mbox{\boldmath $v$}_{\perp}}}
{v_{\perp}^2} , \\
{\mbox{\boldmath $\epsilon$}}_{\perp}&=&
{\mbox{\boldmath $\epsilon$}}
-{\mbox{\boldmath $\epsilon$}}_{\parallel} , \\
t_C&=&\frac{\epsilon_{\parallel}}{v_{\perp}} . 
\end{eqnarray}

For stellar cases in our galaxy, the maximum angular size of 
the distortion is estimated as 
\begin{equation}
\frac{\nu\theta_E}{\beta}\Bigl(\frac{\epsilon}{\beta}\Bigr)^2
\sim \mbox{micro arcsec.} 
\Bigl(\frac{\nu}{10^{-3}}\Bigr)\Bigl(\frac{1}{\beta}\Bigr)
\Bigl(\frac{\theta_E}{\mbox{mas}}\Bigr) 
\Bigl(\frac{\epsilon}{\beta}\Bigr)^2, 
\label{size}
\end{equation}
where we assumed that $\beta$ is comparable to $\epsilon$. 
The maximal distortion occurs at 
\begin{equation}
t_C \sim 10^6 \mbox{s} 
\Bigl(\frac{\epsilon_{\parallel}}{\mbox{AU}}\Bigr)
\Bigl(\frac{100\mbox{km/s}}{v_{\perp}}\Bigr) , 
\label{time}
\end{equation}
about a few months before/after, depending on a location 
of the planet, the source passes the point closest  
to the lensing star. A planet whose orbital separation 
is between 0.1 AU and a dozen of AU will be detectable by future 
missions which observe at intervals from a few days 
to several years. 

\subsection{Discussion}
The position shift falls off as $b^{-1}$, 
while the amplification as $b^{-4}$. 
Hence, the cross section for the astrometric microlensing 
is much larger than that for the photometric microlensing, 
when the sufficient accuracy of astrometric measurements 
will be achieved. 
This is why, up to this point, we have concentrated ourselves 
within a case of a large impact parameter. 
In a small impact parameter case, the caustic crossing 
produces sufficiently large magnification of the images, 
so that we might be able to detect it much more easily 
by the astrometric microlensing as well as photometric microlensing 
(Safizadeh et al. 1999). 
Numerical implementations are needed to study such a nonlinear
behavior. 
In order to evaluate feasibility in the future mission, 
we must take account of the brightness of the lensing star. 
It might be important to pay attention also to fluctuation of 
extragalactic reference frame due to gravitational lensing 
of black holes (Schutz 1982) and MACHOs (Hosokawa et al 1997) 
in our galaxy. 
Finally, the Keplerian motion might cause an appreciable effect on 
the distortion of the photo center ellipse, since the orbital 
period of the planet ranges from months to more than years, 
presumably comparable to $t_C$.  
Therefore, it would be an important subject to study these effects 
in detail. 

\section{Conclusion}
In expectation of dramatic improvements in the precision 
of the future astrometric measurements, we have developed  
a perturbative approach to microlensing 
due to an extrasolar planetary lens. 
Formulae for triple images due to the planetary lens 
are given perturbatively by Eqs. $(\ref{shift1})-(\ref{shift3})$. 
In particular, we have shown by Eqs. $(\ref{size})$ and $(\ref{time})$, 
how the light centroid shifts are dependent on 
the mass ratio and separation between the star and the planet. 
The typical time scale is of the order of months, 
depending strongly on $\epsilon_{\parallel}$, a projection 
of the separation vector onto the source trajectory.

\acknowledgements
The author would like to thank M. Bartelmann, L. Blanchet, 
G. Boerner, C. Cutler, N. Gouda, M. Kasai and B. Schutz 
for fruitful conversation and encouragements.  
He would like to thank G. Boerner, B. Schutz and 
L. Blanchet for hospitality at the Max-Planck Institute for 
Astrophysics, the Albert Einstein Institute and 
the Institute for Astrophysics at Paris, respectively, 
where a part of this work was done. 
This work was supported in part by a Japanese Grant-in-Aid 
for Scientific Research from the Ministry of Education, 
No. 13740137.

\end{document}